%% file: rheograins.tex
\documentclass[twocolumn,showpacs,preprintnumbers,amsmath,amssymb]{revtex4}

\usepackage{graphicx}
\usepackage{bm}

\usepackage{color}

\begin{document}

\title{Dense granular flows: \\ two-particle argument accounts for friction-like constitutive law with threshold}

\author{Pierre Rognon}
\author{Cyprien Gay}

\email{cyprien.gay@univ-paris-diderot.fr}

\affiliation{%
Centre de Recherche Paul Pascal, CNRS UPR~8641 - Av. Dr. Schweitzer, Pessac, France\\
Mati\`{e}re et Syst\`{e}mes Complexes, Universit\'{e} Paris-Diderot - Paris 7, CNRS UMR~7057 - Paris, France
}%

\date{\today}

\begin{abstract}
A scalar constitutive law is obtained for dense granular flows,
both in the inertial regime where the grain inertia dominates,
and in the viscous regime.
Considering a pair of grains rather than a single grain,
the classical arguments yield a constitutive law 
that exhibits a flow threshold expressed as a finite 
effective friction at flow onset.
The value of the threshold is not predicted.
The resulting law seems to be compatible with existing data,
provided the saturation at high velocity (collisional regime)
is added empirically.
The law is not exactly the same in both regimes,
which seems to indicate that there is no ``universal'' law.
\end{abstract}

\maketitle

\newcommand{\hide}[1]{#1}

\newcommand{\hs}{\hspace{0.8cm}}
\newcommand{\dd}{{\rm d}}
\newcommand{\be}{\begin{equation}}
\newcommand{\ee}{\end{equation}}
\newcommand{\bee}{\begin{eqnarray}}
\newcommand{\eee}{\end{eqnarray}}

\newcommand{\transp}[1]{{#1}^T}
\newcommand{\trace}{{\rm tr}}

\newcommand{\gd}{\dot{\gamma}}
\newcommand{\philiq}{\phi}
\newcommand{\phisol}{\psi}
\newcommand{\phisolzero}{\phisol_0}
\newcommand{\visc}{\eta}
\newcommand{\viscapp}{\visc_{\rm app}}
\newcommand{\rayon}{d}
\newcommand{\gapmax}{\Delta}
\newcommand{\masse}{m}
\newcommand{\gap}{h}
\newcommand{\gapmin}{h_{\rm min}}

\newcommand{\contactdensity}{c}
\newcommand{\contactenergy}{{\cal E}_{\rm c}}
\newcommand{\slidingenergy}{{\cal E}_{\rm sl}}
\newcommand{\slidingdistance}{d_{\rm sl}}
\newcommand{\fn}{F_N}
\newcommand{\fnapp}{F_N^{\rm app}}
\newcommand{\fnsep}{F_N^{\rm sep}}
\newcommand{\ft}{F_T}
\newcommand{\vn}{v_N}
\newcommand{\vt}{v_T}
\newcommand{\hmin}{h_{min}}
\newcommand{\coordinance}{z}
\newcommand{\partofdeformation}{f}
\newcommand{\sectioneffective}{A} 
\newcommand{\period}{T}
\newcommand{\periodVIS}{\period^{\rm v}}
\newcommand{\periodINE}{\period^{\rm i}}
\newcommand{\ratio}{\cal {R}}

\newcommand{\Tmicro}{T_{micro}}
\newcommand{\Tsl}{T_{sl}}
\newcommand{\Tapp}{T^{app}}
\newcommand{\Tsep}{T^{sep}}
\newcommand{\Tcar}{T_{car}}
\newcommand{\Tc}{T^\star}
\newcommand{\Tstokes}{T_{Stokes}}
\newcommand{\Tfv}{T_{\rm fv}}
\newcommand{\Tinertiel}{T_{\rm gi}}
\newcommand{\Igi}{I_{\rm gi}}
\newcommand{\Ifv}{I_{\rm fv}}
\newcommand{\Ifi}{I_{\rm fi}}

\newcommand{\si}{\sigma}
\newcommand{\sic}{\sipr^\star}
\newcommand{\sidev}{\tau}
\newcommand{\sipr}{P}

\section{Introduction}

Dense flows of dry granular materials and granular pastes is still a surprising field of research~\cite{Coussot05}. Many complex features have already been observed, but the basic mecanisms that relate grain properties to the collective behavior are not fully identified yet. Gaining new insights into this multi-scale issue should provide robust models and rationalize a wide range of results obtained in various geometries and with various materials. 

Here, we focus on the central question of the constitutive law, which relates the stresses (shear stress $\sidev$ and pressure $\sipr$) to the shear rate $\gd$ in a steady and homogeneous flow. Several measurements evidenced a flow threshold with a frictional criterion: a flow stops if the ratio between the shear stress and the pressure (effective friction coefficient $\mu = \frac{\sidev}{\sipr}$) is lower than a critical friction coefficient $\mu_c$. This is reminicent of yield stress fluids, which stop below a critical shear stress $\sidev_c$. But for granular materials this yield stress depends on the pressure ($\sidev_c = \mu_c \sipr$). Measurements also evidenced an increase of the friction coefficient with the shear rate. 
Furthermore, dimensional analysis implies that the effective friction coefficient cannot depend just on the shear rate $\gd$, but rather on a dimensionless combination $I=T\gd$ where $T$ is some time scale. $T$ is usually understood as the typical time for a single grain, accelerated by the pressure $P$ (or force $Pd^2$), to move over a distance comparable to its own size $d$. It has be estimated in two situations through a simple dimensional analysis. For dry grains, the surrouding fluid can be neglected and the pressure is resisted only by grain inertia (mass $m$)~\cite{Dacruz05a}:

\be
\label{Eq:T_gi}
\masse \, \frac{\rayon}{\Tinertiel^2} \simeq \sipr\,\rayon^2
\hs{i.e.,}\hs
\Tinertiel=\sqrt{\frac{\masse}{\sipr\,\rayon}}.
\ee

\noindent For granular pastes, if the grain inertia is negligible as compared to the viscous effect of the surrounding fluid, the force $Pd^2$ exerced by the pressure is resisted by a typical Stokes viscous drag force felt by a single grain moving at the velocity $d/\Tfv$~\cite{Cassar05}: 

\be
\label{Eq:T_fv}
 \visc\,\rayon\,\frac{\rayon}{\Tfv} \simeq \sipr\,\rayon^2 
\hs{i.e.,}\hs
\Tfv = \frac{\visc}{\sipr}.
\ee

Using these time scales provided robust scaling behaviors in various flow geomerties for dry grains, whether frictional or not~\cite{Dacruz05a,GDR04,Jop06,Forterre08}, cohesive ~\cite{Rognon06a,Rognon08JFM} or polydispersed~\cite{Rognon07PF}, and also for immersed grains~\cite{Cassar05}. Empirical functional form for the constitutive law were proposed to describe the measurements~\cite{GDR04,Dacruz05a,Cassar05,Jop06,Forterre08,Rognon06a,Rognon08JFM,Rognon07PF,Campbell05,Silbert07,Brewster08},for instance~\cite{Cassar05}:
 
\be
\label{Eq:mu_I_cassar}
\frac{\sidev}{\sipr} =\mu(I)=\mu_c+\frac{\mu_2-\mu_c}{I_0/I+1}.
\ee

\noindent This shape includes a critical friction coefficient below which flow stops, $\mu_c$. At large $I$, a saturation of the effective friction ($\mu(I) \rightarrow \mu_2$) reflects the onset of the collisional regime, where grains are diluted and interact through binary collisions~\cite{GDR04,Dacruz05a,Forterre08}. In the dense regime, at small $I$, the effective friction is assumed to increase linearly with $I$ (slope $\frac{\mu_2-\mu_c}{I_0}$), and thus with the shear rate. However, a quadratic term in $\gd$ was also included to fit the data in other works~\cite{Campbell05,Silbert07,Brewster08}. So far, there is no consensus on the shape of the constitutive law. As for the choice of a frictional form like~(\ref{Eq:mu_I_cassar}), it is justified purely phenomenologically. 

In this paper, we propose a simple analysis to derive such a frictional constitutive law from the grain properties. We apply this analysis first to dry grains with no fluid, then to immersed grains with no inertia, and finally to a situation with both the grain inertia and the fluid viscosity. We compare the model thus obtained to the experimental data with immersed and dry grains by Cassar~\textit{et al.}~\cite{Cassar05}, kindly provided by the authors.

Hopefully, the argument presented here may later be useful to understand the origin of much complex features of dense granular flow: the viscosity bifurcations~\cite{Dacruz02,Coussot02c,Coussot06}, a critical shear rate below which no homogeneous flow exists~\cite{Coussot02c, GDR04, Dacruz05a, Cassar05, Rognon08JR} or a possible coexistence of flowing and non-flowing regions (shear-banding, shear localization, cracks)~\cite{0295-5075-66-1-139,Dacruz05a,Huang05,lenoble:073303,Mills08,Rognon08JFM}.

\hide{
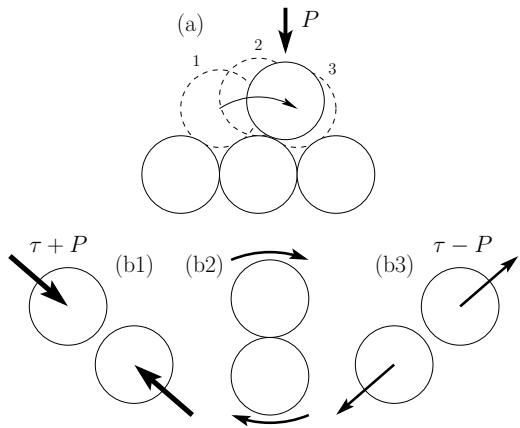
\begin{figure}
\begin{center}
\resizebox{0.8\columnwidth}{!}{%
\input{one_grain_two_grains.pstex_t}
}
\end{center}
\caption{Schematic evolution within the granular material during shear. 
(a) One grain is transported quasistatically
{from} position 1 to position 2, then falls into position 3
due to the applied pressure $\sipr$.
In the present work, we rather consider a pair of grains
during the period of time when they are close neighbours (b1-3).
First, the deviatoric (typically, shear) component of the stress, $\sidev$,
helps the pressure $\sipr$ establish the contact (b1).
Then, the contact rotates due to the overall material deformation (b2).
Finally, the deviatoric stress overcomes the pressure
to break the contact (b3).
Because the pressure is compressive in a non-cohesive granular material,
the typical magnitude of the force transmitted between both grains
is stronger when the contact forms than when it breaks.
}
\label{Fig:one_grain_two_grains}
\end{figure}
}

\section{One-grain {\em versus} two-grain argument}

As mentioned above, the classical approach consists in assuming that the constitutive law
has a frictional form like $\frac{\sidev}{\sipr}=\mu(I)$
and to include a characteristic time such as those provided
by Eqs.~(\ref{Eq:T_gi}) and~(\ref{Eq:T_fv}),
that reflect the motion of a single grain
subjected to the pressure $\sipr$ in the granular material.

As shown on Fig.~\ref{Fig:one_grain_two_grains},
the present approach is focused
on the entire lifetime of a contact between two grains
rather than on the motion of a single grain.
The contact is beeing established when the pair of grains
is oriented along a compressive direction, see drawing (b1).
Hence, the deviatoric part of the stress
is complemented by the pressure part
to favour the approach of both grains.
The opposite is true when the contact is breaking (b3):
the pair is now oriented along a tensile direction,
and the pressure partly resists
the deviatoric part of the stress.
The normal force acting on the contact
is thus stronger during the approach
than during the separation:
\bee
\label{Eq:fnapp}
&\fnapp \simeq \rayon^2\,(\sipr+\sidev)& \\
\label{Eq:fnsep}
&\fnsep \simeq \rayon^2\,(\sipr-\sidev)& \\
&|\fnapp| > |\fnsep|&
\eee
where $\sidev>0$ is the magnitude of the deviatoric stress.
Note that the stress on the contact
evolves as described above
even if the macroscopic stress does not vary.
Indeed it is the rearrangement of the neighbouring grains
that induce the progressive reorientation of the pair of interest
and the corresponding inversion 
of projection of the deviatoric stress on the pair axis.

Let us now assume that during a $100\%$ deformation,
all inter-grain contacts are typically renewed.
The corresponding time $1/\gd$
is then comparable with the typical lifetime of a contact:
\be
\label{Eq:shear_rate_related_to_contact_lifetime}
\frac{1}{\gd} \simeq \Tapp + \Tsep
\ee
where the value of $\Tapp$ and $\Tsep$
is given by Eqs.~(\ref{Eq:T_gi}) and~(\ref{Eq:T_fv}),
replacing the pressure by $\sidev\pm\sipr$
as in Eqs.~(\ref{Eq:fnapp}-\ref{Eq:fnsep}):
\bee
\label{Eq:Tappgi}
\Tapp_{\rm gi} &\simeq& \sqrt{\frac{m}{(\sidev+\sipr)\,\rayon}} \\
\label{Eq:Tsepgi}
\Tsep_{\rm gi} &\simeq& \sqrt{\frac{m}{(\sidev-\sipr)\,\rayon}} \\
\label{Eq:Tappfv}
\Tapp_{\rm fv} &\simeq& \frac{\visc}{\sidev+\sipr} \\
\label{Eq:Tsepfv}
\Tsep_{\rm fv} &\simeq& \frac{\visc}{\sidev-\sipr}
\eee

{From} Eqs.~(\ref{Eq:shear_rate_related_to_contact_lifetime})
and~(\ref{Eq:Tappgi}-\ref{Eq:Tsepgi}),
we obtain a local scalar constitutive equation
for the inertial regime:
\be
\label{Eq:rheo_gi_complete}
\gd\,\sqrt{\frac{\masse}{\sipr\,\rayon}}
=\frac{1}{1+\sqrt{\frac{\sidev-\sipr}{\sidev+\sipr}}}\,
\sqrt{\frac{\sidev}{\sipr}-1}
\ee
which can be approximated as:
\be
\frac{\sidev}{\sipr} \simeq 1+\gd^2\,\frac{\masse}{\sipr\,\rayon}
\label{Eq:rheo_gi_simplified}
\ee
since the first factor in Eq.~(\ref{Eq:rheo_gi_complete})
varies smoothly between $1$ and $1/2$ for $\sidev>\sipr$.

Similarly, using 
Eqs.~(\ref{Eq:shear_rate_related_to_contact_lifetime})
and~(\ref{Eq:Tappfv}-\ref{Eq:Tsepfv}),
we obtain a local scalar constitutive equation
for the viscous regime:
\be
\label{Eq:rheo_fv}
\frac{\sidev}{\sipr}
=\frac{\sipr}{\sidev}+\frac{\gd\,\visc}{\sipr}
\ee

\section{Discussion and comparison with experiments}

When combined with Eq.~(\ref{Eq:T_gi}) 
or~(\ref{Eq:T_fv}) as appropriate,
Eqs.~(\ref{Eq:rheo_gi_complete}) and~(\ref{Eq:rheo_fv}) become:
\bee
\label{Eq:rheo_gi_I}
I &\simeq& \frac{1}{1+\sqrt{\frac{\sidev-\sipr}{\sidev+\sipr}}}\,
\sqrt{\frac{\sidev}{\sipr}-1} \\
\label{Eq:rheo_fv_I}
I &\simeq& \frac{\sidev}{\sipr}-\frac{\sipr}{\sidev}
\eee

Even though these two equations
do not reduce to Eq.~(\ref{Eq:mu_I_cassar}),
let us emphasize that they imply
the existence of a flow threshold:
flow can occur ($\gd>0$)
only when $\sidev>\sipr$.

However, the {\em value} of the threshold,
named $\mu_c$ in Eq.~(\ref{Eq:mu_I_cassar}),
is not predicted in the present approach.
Indeed, the qualitative argument presented earlier,
that leads to a different value for the force
between two particles
during their approach and during their separation
is not precise enough to provide any reliable prefactor
for $\sidev$ in Eqs.~(\ref{Eq:fnapp}-\ref{Eq:fnsep}).

As shown by Fig.~\ref{Fig:fonctions}, 
the saturation of function $\mu(I)$ at large $I$
introduced empirically in Eq.~(\ref{Eq:mu_I_cassar})
is not reproduced by Eqs.~(\ref{Eq:rheo_gi_I}-\ref{Eq:rheo_fv_I}),
as the arguments used to obtain the adimensional parameter $I$
for the dense flows
are not compatible with the binary collisions
thought to be at the origin of the saturation.
However, the plots of Eqs.~(\ref{Eq:rheo_gi_I}-\ref{Eq:rheo_fv_I})
are not very different from that of 
\be
\label{Eq:mu_I_cassar_infty}
\frac{\sidev}{\sipr} - 1 = I
\ee
(see curve C$\infty$ of Fig.~\ref{Fig:fonctions}),
obtained from Eq.~(\ref{Eq:mu_I_cassar}) with $\mu_c=1$,
after taking the limit $\mu_2\rightarrow\infty$
and $I_0\rightarrow\infty$ while keeping $I_0=\mu_2/\mu_c-1$,
in order to remove the saturation.

Let us introduce empirically the very same saturation
into Eqs.~(\ref{Eq:rheo_gi_I}-\ref{Eq:rheo_fv_I})
in order to be able to test 
whether they are compatible with the data 
presented by Cassar {\em et al.}~\cite{Cassar05}.

Eq.~(\ref{Eq:mu_I_cassar}) can be rewritten as:
\be
\label{Eq:mu_I_cassar_saturation}
\frac{\sidev/\mu_c}{\sipr} - 1 =\frac{\mu_2/\mu_c-1}{I_0/I+1}
\ee
Similarly, let us transform 
Eqs.~(\ref{Eq:rheo_gi_I}-\ref{Eq:rheo_fv_I}) into:
\bee
\label{Eq:rheo_gi_I_mu1_I0}
\frac{1}{1+\sqrt{\frac{\sidev/\mu_c-\sipr}{\sidev/\mu_c+\sipr}}}\,
\sqrt{\frac{\sidev/\mu_c}{\sipr}-1} 
&\simeq& \frac{\mu_2/\mu_c-1}{I_0/I+1} \\
\label{Eq:rheo_fv_I_mu1_I0}
\frac{\sidev/\mu_c}{\sipr}-\frac{\sipr}{\sidev/\mu_c}
&\simeq& \frac{\mu_2/\mu_c-1}{I_0/I+1}
\eee

Eqs.~(\ref{Eq:mu_I_cassar_saturation}-\ref{Eq:rheo_fv_I_mu1_I0})
are plotted on Fig.~\ref{Fig:mu_de_I}
in the form $\frac{\sidev}{\sipr}$ as a function of $I$,
together with the data of Fig.~13 from Ref.~\cite{Cassar05},
kindly provided by the authors.

It appears that with adjusted values of $I_0$ and $\mu_2$,
Eqs.~(\ref{Eq:rheo_gi_I_mu1_I0}-\ref{Eq:rheo_fv_I_mu1_I0})
are also compatible with the data.
More elaborate comparisons should be carried out
to see whether the aerial and submarine data
can be usefully distinguished using the above equations.

For completeness, an interpolation
between Eqs.~(\ref{Eq:rheo_gi_I}) and~(\ref{Eq:rheo_fv_I})
has been derived in the same manner
and is available in the form of 
a short note~\cite{Rognon08rheograins_interpolation}.

\hide{
\begin{figure}
\begin{center}
\resizebox{1.0\columnwidth}{!}{%
\input{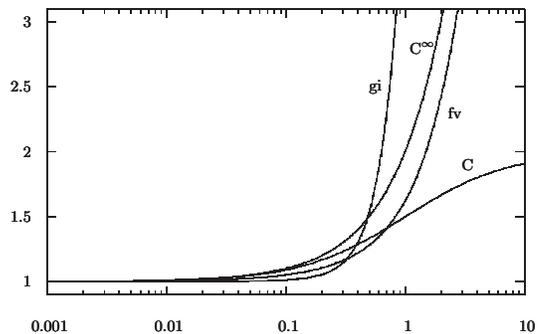}
}
\end{center}
\caption{Comparison of the present predictions
with the empirical function $\mu(I)$.
The curves labeled ``gi'' and ``fv''
are the plots of Eqs.~(\ref{Eq:rheo_gi_I}-\ref{Eq:rheo_fv_I})
respectively.
Curve ``C'' corresponds to Eq.~(\ref{Eq:mu_I_cassar}),
with $\mu_c=1$, $\mu_2=2$ and $I_0=1$.
Curve ``C$^\infty$'' is obtained after removing the saturation,
see Eq.~(\ref{Eq:mu_I_cassar_infty}).
}
\label{Fig:fonctions}
\end{figure}
}

\begin{figure}
\begin{center}
\resizebox{1.0\columnwidth}{!}{%
\includegraphics{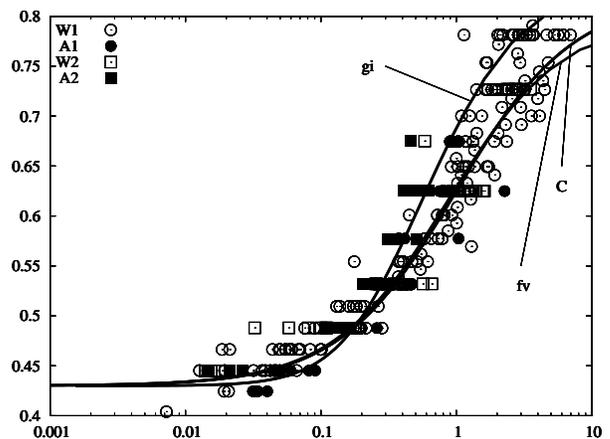}
}
\end{center}
\caption{Effective friction coefficient $\mu(I)$:
data by Cassar {\em et al.}~\cite{Cassar05}
kindly provided by the authors,
and plot of three different functions.
Open circles (W1): 112$\mu$m diameter beads in water.
Filled circles (A1): same beads in air.
Open squares (W2): 208$\mu$m beads in water.
Filled squares (A2): same beads in air.
Curve ``C'' corresponds to Eq.~(\ref{Eq:mu_I_cassar})
with $I_0=1$, $\mu_c=0.43$ and $\mu_2=0.82$.
Curve ``gi'' represents Eq.~(\ref{Eq:rheo_gi_I_mu1_I0})
with $I_0=0.2$, $\mu_c=0.43$ and $\mu_2=0.7$.
Curve ``fv'' corresponds to Eq.~(\ref{Eq:rheo_fv_I_mu1_I0})
with $I_0=0.7$, $\mu_c=0.43$ and $\mu_2=1$.
}
\label{Fig:mu_de_I}
\end{figure}

\section{Conclusion}

Considering a pair of grains rather than a single grain
in a dense granular flow suggests 
that the typical stress during the approach of two grains
differs from that during their seperation.
This difference implies that the usual arguments
used to derive an intrinsic timescale, 
whether when the grain intertia is dominant
or when the fluid viscosity is dominant,
now additionally predict the existence
of a frictional threshold,
with flow only when the ratio
of the deviatoric stress to the pressure
exceeds some finite value: $\sidev/\sipr>\mu_c$.
However, the threshold value $\mu_c$
(usually around $0.4$)
cannot be predicted by this approach.

Cassar {\em et al.}~\cite{Cassar05}
had obtained a ``universal'' behaviour
in the empirical form of Eq.~(\ref{Eq:mu_I_cassar}),
in terms of a dimensionless parameter $I$
with a different meaning in both regimes.
The fact that two distinct laws are found
for the inertial regime and for the viscous regime
with the present approach
indicates that the universal behaviour
is perhaps only approximate.
More involved investigations will be needed
to test this universality.

\subsection*{Acknowledgements}

This work was supported by the Agence Nationale de la Recherche (ANR05).

\bibliography{fluche,./../../Bibliographie}

\end{document}

%% file: one_grain_two_grains.pstex_t
\begin{picture}(0,0)%
\includegraphics{one_grain_two_grains.pstex}%
\end{picture}%
\setlength{\unitlength}{4144sp}%
\begingroup\makeatletter\ifx\SetFigFontNFSS\undefined%
\gdef\SetFigFontNFSS#1#2#3#4#5{%
  \reset@font\fontsize{#1}{#2pt}%
  \fontfamily{#3}\fontseries{#4}\fontshape{#5}%
  \selectfont}%
\fi\endgroup%
\begin{picture}(5982,4888)(1612,-6120)
\put(1891,-4111){\makebox(0,0)[lb]{\smash{{\SetFigFontNFSS{17}{20.4}{\rmdefault}{\mddefault}{\updefault}{\color[rgb]{0,0,0}$\sidev+\sipr$}%
}}}}
\put(6571,-4111){\makebox(0,0)[lb]{\smash{{\SetFigFontNFSS{17}{20.4}{\rmdefault}{\mddefault}{\updefault}{\color[rgb]{0,0,0}$\sidev-\sipr$}%
}}}}
\put(4501,-1771){\makebox(0,0)[lb]{\smash{{\SetFigFontNFSS{12}{14.4}{\rmdefault}{\mddefault}{\updefault}{\color[rgb]{0,0,0}2}%
}}}}
\put(5041,-1501){\makebox(0,0)[lb]{\smash{{\SetFigFontNFSS{17}{20.4}{\rmdefault}{\mddefault}{\updefault}{\color[rgb]{0,0,0}$\sipr$}%
}}}}
\put(3781,-1951){\makebox(0,0)[lb]{\smash{{\SetFigFontNFSS{12}{14.4}{\rmdefault}{\mddefault}{\updefault}{\color[rgb]{0,0,0}1}%
}}}}
\put(3601,-1546){\makebox(0,0)[lb]{\smash{{\SetFigFontNFSS{17}{20.4}{\rmdefault}{\mddefault}{\updefault}{\color[rgb]{0,0,0}(a)}%
}}}}
\put(5896,-4336){\makebox(0,0)[lb]{\smash{{\SetFigFontNFSS{17}{20.4}{\rmdefault}{\mddefault}{\updefault}{\color[rgb]{0,0,0}(b3)}%
}}}}
\put(2881,-4336){\makebox(0,0)[lb]{\smash{{\SetFigFontNFSS{17}{20.4}{\rmdefault}{\mddefault}{\updefault}{\color[rgb]{0,0,0}(b1)}%
}}}}
\put(3691,-4336){\makebox(0,0)[lb]{\smash{{\SetFigFontNFSS{17}{20.4}{\rmdefault}{\mddefault}{\updefault}{\color[rgb]{0,0,0}(b2)}%
}}}}
\put(5356,-2041){\makebox(0,0)[lb]{\smash{{\SetFigFontNFSS{12}{14.4}{\rmdefault}{\mddefault}{\updefault}{\color[rgb]{0,0,0}3}%
}}}}
\end{picture}%